\documentclass[mathleft
]{an}
\usepackage{graphicx}
\usepackage{times}
\begin{document}

\Pagespan{789}{}
\Yearpublication{2006}%
\Yearsubmission{2005}%
\Month{11}%
\Volume{999}%
\Issue{88}%

\title{Black-hole states in external galaxies}

\author{Tomaso M. Belloni\inst{1}\fnmsep\thanks{Corresponding author:
  \email{tomaso.belloni@brera.inaf.it}\newline}
}
\titlerunning{Black-hole states in external galaxies}
\authorrunning{T.M. Belloni}
\institute{
INAF - Osservatorio Astronomico di Brera,
Via E. Bianchi 46,
I-23807 Merate, Italy
}

\received{?? Sep 2010}
\accepted{?? ??? 201?}
\publonline{later}


\abstract{%
A large number of Ultra-Luminous X-ray Sources in external galaxies is now known, but the discussion about their nature is still unsettled: intermediate-mass black holes or stellar-mass black holes accreting above the Eddington limit? A promising path that can provide an answer is through comparison with ``normal'' stellar-mass black holes. These display separate states, identified from both their energy spectra and their fast variability. Although the nature of these states and their transitions is not clear, such a comparison can be extremely useful. In this paper, I briefly outline our current phenomenological knowledge of these states and present a (partial) overview of the different approaches followed to make these comparisons.
}

\maketitle

\section{Introduction}

The nature of Ultra-Luminous X-ray Sources (ULX) in external galaxies is still under debate. Since their 
observed luminosity exceeds the Eddington limit for a stellar-mass black hole, the three possible interpretations are: (1) their mass is much larger than 10$M_\odot$ or (2) their emission is not isotropic, i.e. beamed towards us (see other articles in this issue for an overview) or (3) they emit above the Eddington luminosity. 

While the observational properties of ULXs do not allow yet to constrain the mass of the central object, it is reasonable to expect that accretion in these objects can be compared to that onto super-massive and stellar-mass black holes. In particular, stellar-mass black holes offer the clearest phenomenology of variability which takes place on time scales accessible for 100-1000 $M_\odot$ counterparts.
The main problems with this approach are two. First, the distance to external galaxies results in lower statistics, currently limiting the possibility of a detailed comparison. Second, major aspects of accretion onto X-ray binaries are not yet understood in terms of physical models, in particular when dealing with fast aperiodic variability. This means that the comparison will often be phenomenological, although this in turn will offer a way to obtain a better understanding of the accretion process at large.

In this paper, I outline the current observational status on black-hole binaries in terms of source states and examine the recent results on ULXs to give an overview of the level of comparison that can be made at present together with the possibilities for the near future.

\section{States in black-hole binaries}

Thanks in particular to the Rossi X-Ray Timing Explorer (RXTE), the past decade has seen a large increase in the observational information on galactic black-hole binaries (BHB), in particular on transient systems. While it is possible to extract and analyze full energy spectra at low (e. g. RXTE) and high (e.g. Chandra, XMM-Newton) resolution, the complexity of these observables makes a classification very difficult. The time evolution of transients is very varied and offers little information common to all sources and outbursts, unlike the pre-RXTE situation when most of the handful of known transient sources showed a rather simple fast rise-exponential evolution (with complications: see Tanaka \& Lewin 1996). One thing which is important to stress, as it is often overlooked, is that any robust state classification must necessarily be based both on spectral and timing properties (see Remillard \& McClintock 2006; Belloni 2010).

However, using crude X-ray estimators allows to reduce the complex variations within a simple scheme which provides a framework to which complex models can be applied and tested (see Belloni 2010). Three (instrument dependent) quantities are easy to extract from the data: the total source count rate within a reasonably broad band, the hardness-ratio (the ratio of hard to soft photons) and the amount of variability (whether expressed in absolute or fractional terms). They vary throughout the outburst of a transient (but also in persistent systems) and their combination defines precisely the source state.
In the context of ULX comparison, the most useful tool is the Hardness-Intensity Diagram (HID), where net source count rate is plotted as a function of spectral hardness (defined as the ratio of counts in two energy bands, with the higher-energy band at the numerator).

The position of a source in the HID is determined by its energy spectrum in the 3-20 keV energy band (from the RXTE/PCA). We know that this spectrum is rather complex (see, e.g., Gilfanov, 2010), but the two major components which are (or can be) present are a thermal component, interpreted as the emission from a thermal optically-thick and geometrically-thin accretion disk, plus a harder component which in this energy range can roughly be described by a power law with large variations in its spectral index. 
Once all observations from an outburst of a transient or all observations of a persistent system are put on this diagram, a pattern appears. First, the source populates well-defined regions of this diagram. Second, many sources follow a similar pattern in their time evolution throughout the outburst. The evolution of the energy spectrum, however, is not sufficient to characterize the ``state" of the source. In addition, the properties of the time variability on time scales shorter than $\sim$10 seconds have to be considered. The analysis of power density spectra and phase lag spectra shows different components such as noise and Quasi Periodic Oscillations (QPO). These components show abrupt changes which cannot be ignored for a classification in terms of discrete states (see Belloni et al. 2005; Belloni 2010).

A sketch of the HID can be seen in Fig. \ref{fig:hid}. There are two ways of looking at this diagram: a dynamic and a static one. Dynamically, one can follow the evolution of one source (or one outburst) in the diagram. A number of sources (see Homan \& Belloni 2005; Fender, Homan \& Belloni 2009; Belloni 2010) follow roughly a `q' shaped path on the diagram in a counterclockwise direction (see Fig. 1). 
Independent of the dynamical evolution throughout outbursts, less interesting for the case of ULX, one can adopt a static view and examine the spectral and timing properties associated to different regions (or branches) in the HID. In particular, when the fast time variability is considered a number of well-defined states can be identified (see Fig. 2). These states are: Low Hard State  (LHS), Hard Intermediate State (HIMS), Soft Intermediate State (SIMS) and High Soft Sta\-te (HSS). More physics-based diagrams such as the Disk Fraction Luminosity Diagram (DFLD, K\"ording, Jester \& Fender 2006; Dunn et al. 2010) can also be used.

\begin{figure} 
   \includegraphics[width=8.36cm]{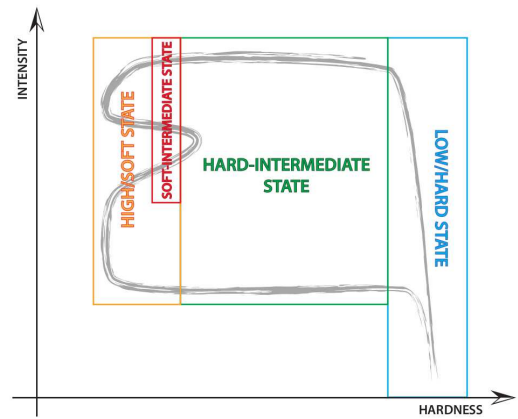} 
   \caption{Sketch of a Hardness-Intensity Diagram (HID) for a transient black-hole binary with the areas corresponding to the different states marked.} 
   \label{fig:hid} 
\end{figure}

The time variability of these states is markedly different. While the LHS is characterized by a very strong (20-40\% fractional rms) aperiodic noise as the sum of broad-band components in its Power Density Spectrum (PDS), the HSS displays very little variability (a few \% fractional rms), also in the form of continuum components in the PDS. The main feature of the intermediate states is the presence of strong quasi-periodic oscillations (QPO). These are of four types: A, B, C (all at low frequencies) and high-frequency (HF) (see Casella, Belloni \& Stella 2005; Belloni et al. 2005; Motta et al. 2010 in prep.). As type-A QPOs are very weak and elusive even for BHBs, they can be ignored here. Type-C QPOs are the most common, found in the HIMS: they are associated to band-limited noise, their frequency ranges between 0.01 and 20 Hz for a single source, evolving smoothly throughout the state (softer spectra have higher frequency). There is a good correlation between their frequency and the spectral index of the hard spectral component (Vignarca et al. 2003). A harmonic series of peaks is usually observed.
Their fractional rms is typically around 10\%.
Type-B QPOs, found in the SIMS (which they define), are associated to weak power-law noise and span a much more limited range in frequency (1-6 Hz), correlated with source flux. Their fractional rms is a few \%. Harmonic peaks are often present.
Figure \ref{fig:qpo} shows two cases QPOs of B and C types, chosen to have the same centroid frequency. The main difference between these two is the presence/absence of broad-band noise (other differences are less evident in this plot, see Belloni 2010).
Finally, HF QPOs are very rare, of a few \% fractional rms, with frequencies between 70 and 450 Hz, found in the SIMS. Of the handful of existing detections, in few a second QPO is present. The frequency of the higher of the two QPOs scales roughly with the black-hole mass.

\begin{figure} 
   \includegraphics[width=8.36cm]{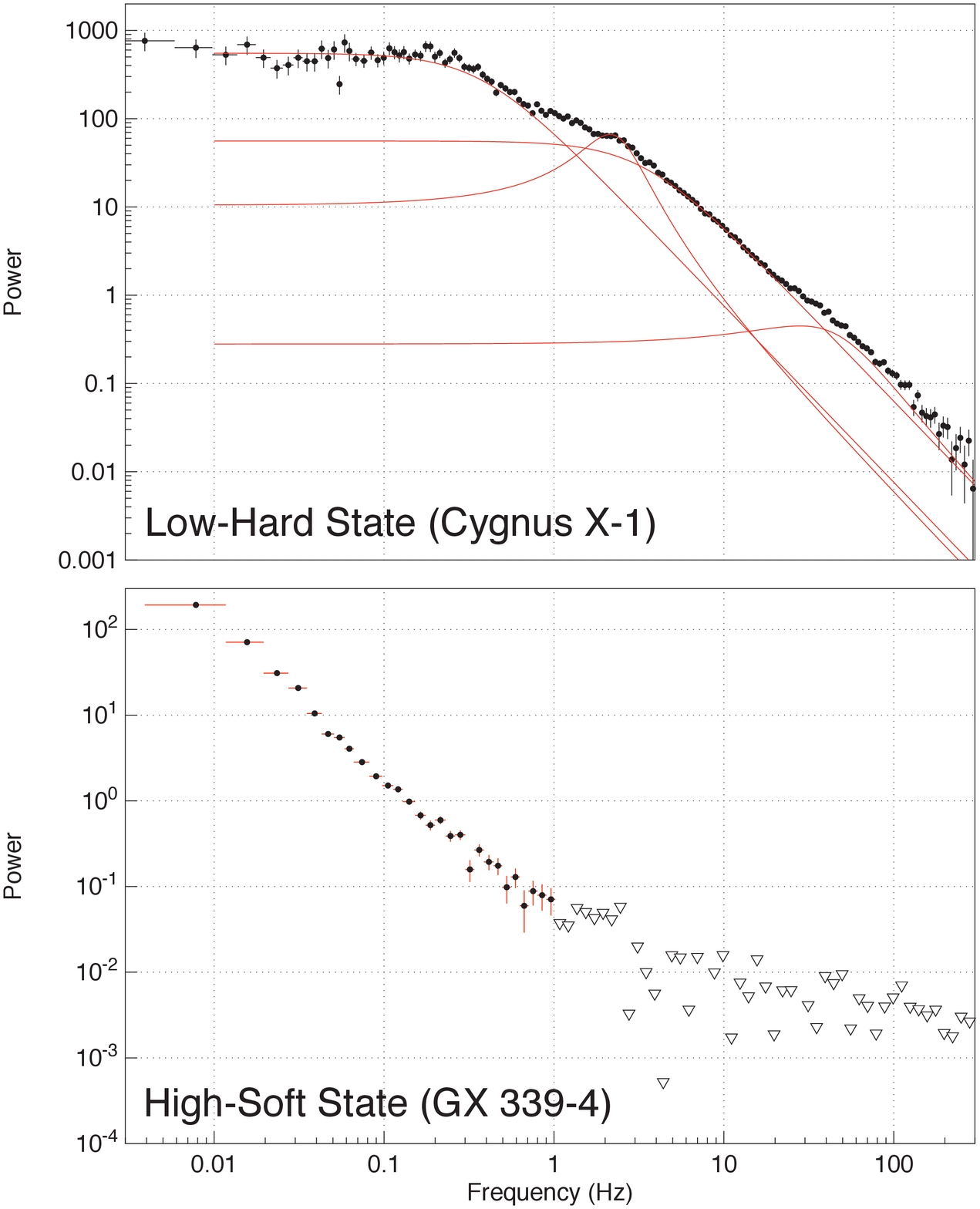} 
   \caption{Top panel: a LHS power spectrum of Cygnus X-1. The total fractional rms is around 30\%. The red lines are the fit components.
   Bottom panel: a HSS power spectrum from a 57-ks RXTE observation of GX 339-4. The total fractional rms is a few \%} 
   \label{fig:hid} 
\end{figure}

The most common states, where sources spend most of their time, are the LHS and the HSS, a discussion whose characteristics can be found in Gilfanov (2010). A comparison on the basis of these two states is useful, but it relies mostly on detailed spectral fitting, which is difficult for BHB due to high signal-to-noise and equally difficult for ULX due to low signal-to-noise. The time variability is also complicated due to low statistics.
The two intermediate states are particularly interesting as they are associated with QPOs. The centroid frequency of a QPO, a very robust measurement, can be compared between BHB and ULX and, depending on the adopted model, a scaling law can yield the mass of the compact object in the latter. The problem here lies in the absence of a general physical model to interpret QPOs.

\begin{figure} 
   \includegraphics[width=8.16cm]{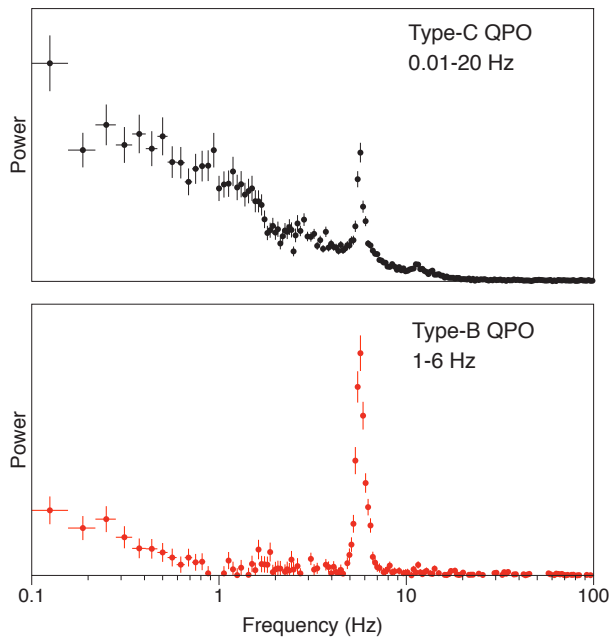} 
   \caption{Top panel: a type-C QPO at 6 Hz (with broad-band noise). Bottom panel: a type-B QPO at the same frequency for comparison. Broad-band noise here is replaced by weaker power-law noise.} 
   \label{fig:qpo} 
\end{figure}

\section{States in ULX}
Two separate paths to compare BHB and ULX can be followed: one based on energy spectra and one based on variability. 
\subsection{Spectral approach}

Fitting energy spectra of ULXs to the same models used for BHBs allow to compare directly their emission. The advantage is that we have physical models for the origin of the emission (although not 100\% understood). For instance, a thermal accretion disk model can in principle provide a measurement of the mass of the black hole through the innermost disk radius. LHS-type and HSS-type spectra have been reported for ULX in the past years.
Interesting are the intermediate cases, i.e. energy spectra which are neither disk nor hard-component dominated. An example is the two ULX in NGC~1313, for which Miller et al. (2003) found evidence of double-component spectra. The presence of a hard component together with a disk component with inner temperature $\sim$150 eV as opposed to the $\sim$1 keV found in intermediate states of BHB suggests a high mass (100-1000 M$_\odot$).

BHBs undergo state transitions. While the transition itself is very fast and difficult to observe in ULXs with only a few observations, the observation of two separate states at different times can also be compared with galactic systems. Holmberg II X-1 was observed in a bright state in April 2002 and then in a low-flux (by a factor of four) state five months later (Dewangan et al. 2004). However, in the two-component spectrum fitted to the XMM-Newton data the fainter spectrum corresponded to the softer spectrum.
This is not inconsistent with BHBs, as both the soft and the hard state can be found over a rather large (and overlapping) range in luminosities.

More observations of the same system can help the spectral analysis and make the results more robust. This is an area where interesting results are starting to appear. The analysis of a few observations of NGC 5204 X-1 with XMM and Chandra showed an evolution similar to that observed in BHBs: brighter spectra show a steeper hard component and a hotter inner disk which follows the $\propto T^4$ law (Feng \& Kaaret 2009). However, the same authors find that the same is not true for IC 342 X-1. Later, the analysis of a source in M 82 which was not classified as an ULX showed that it had moved to an ultra-luminous state, again with the expected temperature changes (Jin, Feng \& Kaaret 2010).
Other observations lead to varied conclusions. The analysis of a large number (12) of XMM-Newton observation of NGC 1313 led to the conclusion that NGC 1313 X-1 showed spectral variations similar to those of galactic systems from LHS to intermediate states, but never reaching the soft state (Feng \& Kaaret 2006). On the other hand, the same work reports that NGC 1313 X-2 is incompatible with that picture, showing a LHS spectrum that hardens as the source brightens, something never observed in galactic sources. The same NGC 1313 X-2 was subsequently found to have also a luminosity-temperature dependence opposite to what expected but compatible with a super-Eddington model (Feng \& Kaaret 2007a).
Recently, evidence for two separate states was presented for NGC 1313 X-1 by Dewangan et al. (2010). Besides a clear spectral difference, large ($\sim$ 15\%) variability was observed connected to the hard spectrum and low ($<$3\%)  with the soft spectrum, in line with the Galactic counterparts. 
Intensive monitoring with Swift is also very useful to follow spectral variability. Kaaret \& Feng (2009) monitored four systems and discovered a strong brightening of the nuclear source in NGC 4395. The spectrum of Hol IX X-1 was followed with Swift: subtle spectral variations of a Comptonized component were interpreted in the framework of a corona whose density increases as the source brightens, consistent with material being blown away from a super-Eddington disk (Vierdayanti et al. 2010).

\begin{figure} 
   \includegraphics[width=8.16cm]{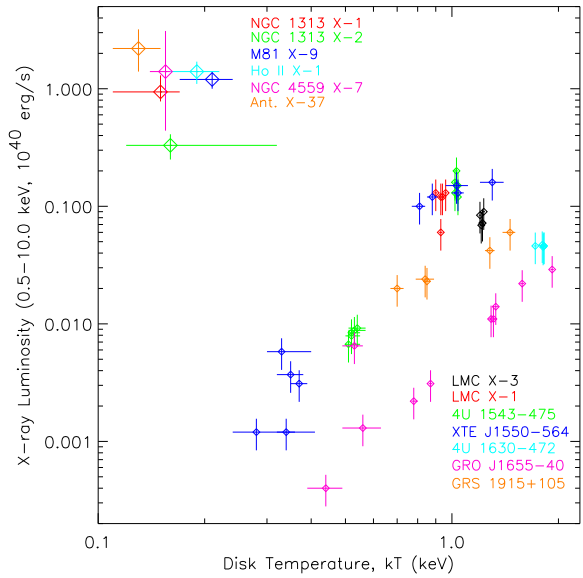} 
   \caption{X-ray luminosity versus disk inner temperature inferred from X-ray spectral fits for a sample of ULXs and of BHBs. From Miller, Fabian \& Miller (2004).} 
   \label{fig:miller} 
\end{figure}

\begin{figure} 
   \includegraphics[width=8.16cm]{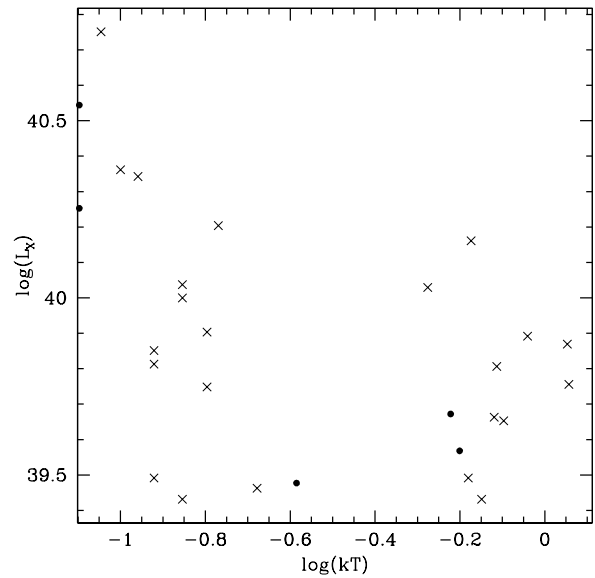} 
   \caption{Blackbody temperature versus 0.3--10 keV luminosity for HSS objects (from Winter, Mushotsky \& Reynolds 2006). The presence of two population is apparent.} 
   \label{fig:winter} 
\end{figure}

Very promising are the studies on samples of sources, although they of course assume that all sources belong to the same class, while it is possible, if not likely, that not all ULXs are the same type of system. Miller, Fabian \& Miller (2004) compared a sample of ULXs and BHBs in a L$_X$ vs. $kT_{in}$ plane (see Fig. \ref{fig:miller}).
Clearly the ULXs occupy a different region of the plane, corresponding to a much larger inner radius, which in turn means higher mass. Although this analysis is complicated by the presence of galaxy emission, this plot is the first approach to source samples.
Winter, Mushotsky \& Reynolds (2006) analyzed a large sample of XMM-Newton ULXs and find 16 consistent with a LHS and 26 with a HSS. The inner disk temperatures of the HSS objects were found to belong to two separate distributions: one around 0.1 keV and one around 1 keV(Fig. \ref{fig:winter}). The former were identified with intermediate-mass objects, the latter with stellar-mass objects. The sample also yielded a set of spectral parameters which are important for comparison. 
Complex spectral results were presented by Kajava \& Poutanen (2009) from the analysis of a sample of 11 systems. They found sources with a power-law spectrum where the spectral index correlated positively with luminosity, as in BHB, and a couple where there is an anticorrelation. For the sources that can be fitted with a single thermal component, the $kT$ vs. $L_X$ correlation is the expected $T^4$. The power-law fits can be improved also by the addition of a soft component with a different $kT$ vs. $L_X$ relation. The authors discuss the possibility that the soft flux could be disk emission beamed by an outflowing wind.

\begin{figure} 
   \includegraphics[width=8.16cm]{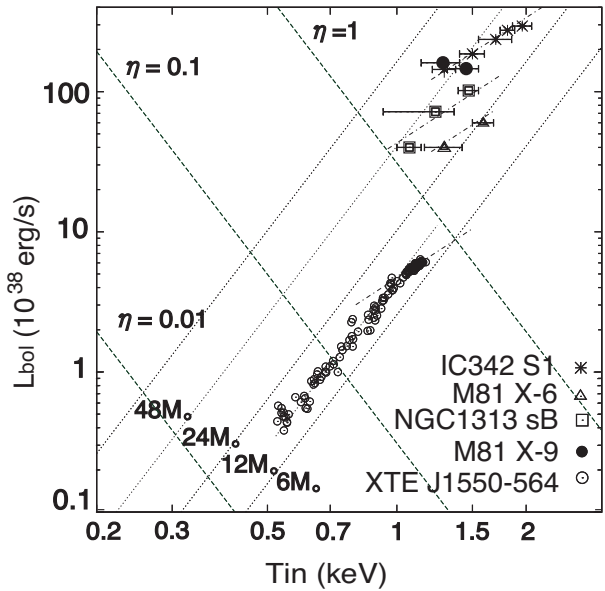} 
   \caption{Blackbody temperature versus bolometric luminosity for multicolor-disk fits to ULXs and bright BHBs. The dotted and dot-dashed lines represent $L_{bol} \propto T^4$ and $L_{bol} \propto T^2$ respectively. The dashed lines are $L_{bol} \propto T^{-4}$ (corresponding to constant Eddington ratio $\eta = L_{bol}/L_{Edd}$). From Tsunoda et al. (2006).} 
   \label{fig:tsunoda} 
\end{figure}

Models beyond standard accretion disk models are also being investigated. They were developed for BHB to interpret the peculiar spectra at very high luminosity (see e.g. Middleton et al. 2006 for energy spectra and Belloni 2010 for time variability). The presence of an ``ultraluminous'' or ``anomalous'' state has been proposed by Tsunoda et al. (2006) and Gladstone et al. (2009). Figure \ref{fig:tsunoda} shows the $L_{bol}$ vs. $kT$ relation for BHBs at high accretion rate and a sample of ULXs. The deviation from the $T^4$ curve at high luminosity is apparent.

\subsection{Timing approach}

A high count rate is needed for the analysis of aperiodic time variability, which means that results can be obtained only for few sources in external galaxies with current instruments. As a result, there are very few systematic studies and the field is mostly made of single detections of timing features.
Heil, Vaughan \& Roberts (2009) analyzed 19 XMM-Newton observations of ULXs: they found six objects with significant variability and four with a stringent upper limit well below those and also below the corresponding variability observed in galactic sources. They conclude that in some ULXs variability is suppressed, although this effect could be due to observational effects. Once a larger database of observations is available, this type of study will be crucial for the comparison between classes of systems.
The same group reported the observation of a linear rms-flux relation in NGC 5408 X-1 and the first measurement of time delays between hard and soft photons (Heil \& Vaughan 2010) The rms-flux relation is  similar to that observed in galactic BHBs and AGN, but the time delays are reversed.
Still, important measurements such as the hardness-rms diagram for a sample of sources have not been made.

\begin{figure} 
   \includegraphics[width=8.16cm]{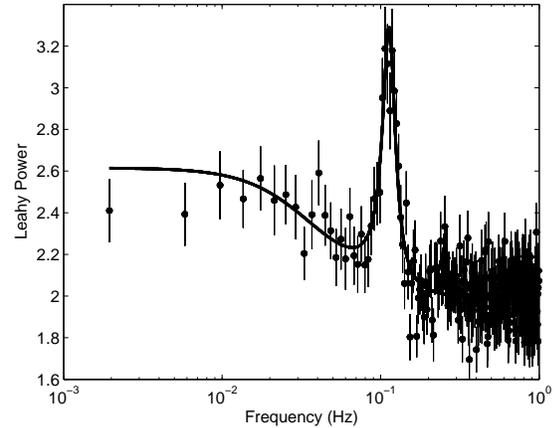} 
   \caption{Total power spectrum of M~82 X-1 from the 2004 XMM-Newton observation (from Mucciarelli et al. 2006).
   } 
   \label{fig:m82qpo} 
\end{figure}

Strohmayer \& Mushotsky (2003) discovered a QPO in an XMM-Newton observation of M~82 X-1. Its centroid frequency was 54.5 mHz with a FWHM of $\sim$11 mHz and an integrated fractional rms of 8.4\% in the 20--10 keV band. They also found evidence in RXTE data for another (non-simultaneous) peak at 107 mHz.  Because of the point spread function of XMM-Newton, it was debated whether the QPO comes indeed from X-1 or from a nearby source (Feng \& Kaaret 2007b).
A later long XMM-Newton observation of M~82 X-1 showed the QPO to be still present, but stronger and at a different frequency (see Fig. \ref{fig:m82qpo}, Mucciarelli et al. 2006). The fractional rms was 18.3\% and the centroid frequency 113 mHz, around twice the previous value. The signal was stronger than in 2002 and it was possible to look for variability within the observation: the QPO drifted by 20\% during the observation.
In addition to the QPO, a noise component is apparent in Fig. \ref{fig:m82qpo}, with a characteristic frequency of $\sim$40 Hz and a fractional rms of 22\%.
Long-term variability and the presence of a strong noise component led the authors to identify the QPO as of type C. Unfortunately, that QPO was observed in BHBs over a large range of frequencies, even lower than the one observed here. In principle, this detection would allow for a 1$M_\odot$ black hole.
Moreover, the QPO was also observed in a series of RXTE observations only at frequencies compatible with 0.5 or 1.5 times the second XMM-Newton detection, making it appear at frequencies in 1:2:3 ratios, something not observed in galactic counterparts.

Another  QPO was discovered with XMM-Newton from NGC 5408 X-1 (Strohmayer et al. 2007) at a frequency of 20 mHz, with a quality factor ($\nu_0/\Delta\nu$) of 6 and a fractional rms of 9\%. Also in this case a strong broad component is observed, with a break around the QPO frequency. Selecting when the QPO is more prominent, there is evidence for a second QPO peak around 15 mHz. Again, this is similar to type-C QPOs, aside for the second non-harmonic peak.
A second XMM-Newton observation of NGC 5408 X-1showed the QPO at 10 mHz (Strohmayer \& Mushotsky 2009). A comparison with the previous observations shows that the QPO frequency increases with disk flux, while its rms (and that of the noise) decreases, in analogy of what seen in BHBs. Using the known correlation between QPO frequency and power-law spectral index for BHBs (see Vignarca et al. 2003) they derive a mass estimate of 2000-5000 $M_\odot$.

\begin{figure} 
   \includegraphics[width=8.16cm]{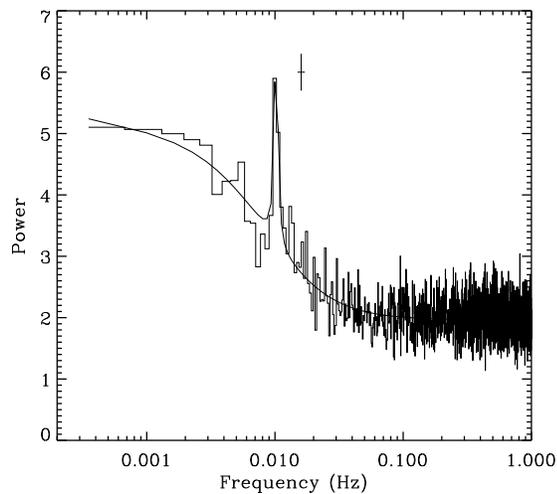} 
   \caption{Power spectrum of NGC 5408 X-1 in the second XMM-Newton observation (from Strohmayer \& Mushotsky 2009). A characteristic error bar is shown.
   } 
   \label{fig:5408} 
\end{figure}

An approach based on general correlations between parameters was followed by Casella et al. (2008). We observe clear QPOs in two systems. In galactic BHBs, the QPO frequency shows a very tight correlation over two orders of magnitude with a higher-frequency feature (Psaltis, Belloni \& van der Klis 1999; Belloni, Psaltis \& van der Klis 2002). The frequency of this feature is known to follow a correlation over eight orders of magnitude with mass accretion rate encompassing BHBs and AGN, after correction for the black-hole mass (K\"ording et al. 2007). Therefore, knowing the QPO frequency and the mass accretion rate the combination of these correlations can give us an estimate of the black hole mass. In order to extract the accretion rate from the observed luminosity one needs to assume an efficiency. Applying this indirect method to the two QPO detections and assuming conservative errors in the correlations, the authors obtain masses in the range 100-1300 $M_\odot$.

\section{Conclusions}

The comparison between X-ray properties of galactic X-ray binaries and Ultra-luminous X-ray sources in external galaxies appears to be one of the most promising approaches to unveil the nature of the latter. While they are probably not a homogeneous population, similar spectral and timing properties have been found for some (bright) systems. Longer and extended campaigns with sensitive instruments such as XMM-Newton an Chandra are needed to enlarge the existing sample of detailed properties, through both new observation of fainter systems and repeated observations of bright ones.

\acknowledgements
The author acknowledges support from ASI grant I/088/06/0 and from the the European CommunityÕs Seventh Framework Programme (FP7/2007-2013) under grant agreement number ITN 215212 ÔBlack Hole UniverseÕ.
I also thank the anonymous referee for his/her constructive comments.

\newpage

\end{document}